# Large and Controllable Spin-Valley Splitting in Two-Dimensional WS$_2$/*h*-VN Heterostructure

Congming Ke[1], Yaping Wu*[1], Weihuang Yang[2], Zhiming Wu[1], Chunmiao Zhang[1], Xu Li[1], Junyong Kang*[1]

[1]Department of Physics, OSED, Fujian Provincial Key Laboratory of Semiconductor Materials and Applications, Jiujiang Research Insititute, Xiamen University, Xiamen, 361005, P. R. China.

[2]Key Laboratory of RF Circuits and System of Ministry of Education, Hangzhou Dianzi University, Hangzhou 310018, P. R. China.

* Yaping Wu (email: ypwu@xmu.edu.cn), and Junyong Kang (email: jykang@xmu.edu.cn). Tel.: +86-592-2181366; Fax: +86-592-2187737.

Inspired by the profound physical connotations and potential application prospects of the valleytronics, we design a two-dimensional (2D) WS$_2$/*h*-VN magnetic van der Waals (vdW) heterostructure and study the control of valley degree of freedom through the first-principles calculations. A considerable spin splitting of 627 meV is obtained at the *K* valley, accompanied with a strong suppression of that at the *K'* valley. An intrinsic large valley splitting of 376 meV is generated in the valence band, which corresponds to an effective Zeeman magnetic field of 2703 T. Besides of the valence band, the conduction band of WS$_2$ possesses a remarkable spin splitting also, and valley labelled dark exciton states are present at the *K'* valley. The strengths of spin and valley splitting relied on the interfacial orbital hybridization are further tuned continually by the in-plane strain and interlayer spacing. Maximum spin and valley splitting of 654 and 412 meV are finally achieved, respectively, and the effective Zeeman magnetic field can be enhanced to 2989T with a -3% strain. Time-reversal symmetry breaking and the sizable Berry curvature in the heterostructure lead to a prominent anomalous Hall conductivity at the *K* and *K'* valleys. Based on these finding, a prototype filter device for both the valley and spin is proposed.

## I. INTRODUCTION

Valleytronics in 2D materials is of fertile ground for fundamental science as well as of great practical interest toward the seamless integration of information processing and storage.[1-4] Valleys, which label the degenerate energy extreme of conduction band or valence band at special *k* points (*K* and *K'*), have large separation in the momentum space that enable valley pseudospin very robust against phonon and impurity scatterings.[5-7] In order to distinguish and manipulate the carriers at defined valleys, the *K* and *K'* valley degeneracy should be lifted to introduce the valley polarization. Plenty of exotic properties, such as quantum spin/valley anomalous Hall effect,[8,9] valley dependent optoelectronics,[10,11] magneto-optical conductivity[12] and electrical transport of valley carriers[13-15] have been explored in valley-polarized systems. Searching feasible approaches to realize valley polarization is of fundamental importance.

Breaking the balance of carriers in the inequivalent valleys is necessary to realize valley polarization, and previous studies have provided various strategies.[7,16-25] One was using an ultrafast circularly polarized laser pump to break the valley degeneracy through the optical Stark effect,[16,17] another was applying a vertical magnetic field by taking advantage of the Zeeman effect.[18-20] However, optical pumping is not suitable for valleytronic applications owing to the difficulty in control, and the efficiency of external magnetic field is usually too low that 1T magnetic field can only give rise to a splitting of 0.1-0.2 meV. Some scientists tried to achieve the valley polarization in manganese chalcogenophosphates by coupling the valley degree of freedom to the antiferromagnetic order, and manipulated the splitting by doping.[21,22] Unfortunately, this way turned out to be rather modest. It was found that considerable valley splitting can be achieved through magnetic proximity coupling by constructing heterostructures with magnetic substrates such as EuO,[23] CoO[24], MnO[7] and EuS,[25] but the bulk substrates intrinsically limit the device applications in nanoscale. Moreover, owing to the polycrystalline nature and small grain size of the 3D bulk substrates, additional magnetic field is required to polarize the ferromagnets for any observable magnetic functionality. Beyond all above methods, constructing perpendicular 2D magnetic vdW heterostructures by layered 2D materials can be a more advantageous mean. A 2D magnetic vdW heterostructure is favored for forming a relatively clean interface to eliminate the impurity scattering, and can minimize the effect of lattice mismatch that would weaken the valley splitting.[26] Additionally, heterostructures constructed completely by 2D materials are more easily to be integrated when fabrication valleytronic devices.[27] In light of these advantages, design 2D layered magnetic/semiconductor vdW heterostructures and investigating the spin-valley splitting are of great significance. In transition-metal dichalcogenides (TMDs; $MX_2$, where M = Mo and W, and X = S and Se) monolayer, inversion symmetry for space breaking together with strong spin-orbit coupling provides an outstanding foundation in valleytronics.[7] On the other hand, h-VN monolayer was predicted to be a ferromagnetic half-metallic material with a high Curie temperature of 768 K, which can be used for spin injection and valley polarization for TMDs monolayer.[28] In the meantime, considering the diminutive lattice mismatch between $WS_2$ and h-VN monolayer (3.19 Å for $WS_2$[7] and 3.23Å for h-VN[28]), constructing $WS_2$/h-VN heterostructures may give rise to novel valleytronic properties and predict promising applications in valleytronic devices.

In this work, we study the control of valley degree of freedom in monolayer $WS_2$/h-VN heterostructure through the first-principles calculations. Both spin and valley properties are calculated and tuned by in-plane strain and interlayer spacing. Large spin and valley splitting are produced through the interfacial orbital hybridization between $WS_2$ and h-VN. The Berry curvature at the K and K' valleys shows opposite signs, and nonzero anomalous Hall conductivity when shifting the Fermi level between the K valley and Γ point is demonstrated. The findings in the work indicate potential applications of 2D magnetic vdW heterostructure in valleytronic devices.

## II. THEORY AND COMPUTATIONAL METHOD

First-principles density functional calculations are performed using the projector augmented wave (PAW) method implemented in the Vienna ab initio simulation package (VASP).[29-31] The generalized gradient approximation (GGA) of Perdew-Burke-Ernzerhof (PBE) parametrization[32] is used for the exchange-correlation potential and the vdW interaction is taken into account using the DFT-D2 method.[33] The fully relativistic projector augmented potential is adopted in order to include the spin-orbit coupling. The valence configurations of V, N, W and S atoms considered in the calculations are $3p^63d^34s^2$, $2s^22p^3$, $5p^65d^46s^2$ and $3s^23p^4$, respectively, in which the d-orbitals of cations are included as semi-core states aimed to improve the computational accuracy. The slab-supercell approach is adopted, and a 25 Å thick vacuum along the z direction is constructed. A large planewave energy cutoff of 520 eV is used, and the Brillouin zone is sampled with a 19 × 19 × 1 Monkhorst-Pack grid of k points. All atomic degrees of freedom, including lattice constants are fully relaxed with self-consistent convergence criteria of 0.01 eV/Å and $10^{-6}$ eV for the atomic forces and total energy, respectively. In the calculations of Berry curvature and anomalous Hall conductivity, the maximally localized Wannier functions (MLWFs)[34] as implemented in the WANNIER90 package[35] are employed. All orbitals of V, N, W and S atoms are selected as the initial orbital projections, and a finer 21 × 21 × 1 uniform k grid is used for the construction of the maximally localized Wannier functions. The difference in the spread of total Wannier functions between two successive iterations converges to $10^{-10}$ Å² within 300 iterative steps.

## III. RESULTS AND DISCUSSION

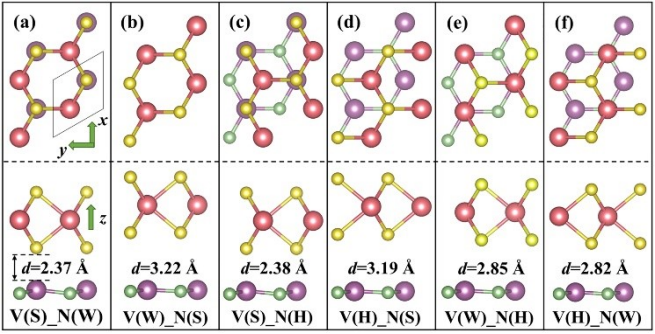

**Fig. 1** (a) Configurations of the WS$_2$/h-VN heterostructures, where the rhombus indicates a unit cell. The red, yellow, pink and green balls correspond to the W, S, V and N atoms, respectively. The related atomic positions of the WS$_2$ atoms to the h-VN atoms (either V or N) are indicated in the brackets, where the H denotes the location of the V or N atoms under the hollow site of WS$_2$ hexagons.

Monolayer WS$_2$ we used has a 2H structure with $P6_3/mmc$ space group, and h-VN monolayer is assembled of its bulk (111) plane with $P\bar{6}2m$ space group. The unit cell of h-VN consists of one V and one N atom in its planar honeycomb lattice, in which each V atom is three-coordinated with N atoms (h-BN like structure). The optimized WS$_2$ and h-VN monolayers have the lattice constants of 3.19 and 3.23 Å, respectively, with a lattice mismatch of 1.2%. According to the symmetry, six kinds of stacking configurations as illustrated in Fig. 1 are taken into account for the WS$_2$/h-VN heterostructures. All the configurations are fully relaxed respectively to optimize the interlayer spacing between WS$_2$ and h-VN monolayers. To determine the most stable configuration, binding energies $E_b$ between the WS$_2$ and h-VN monolayer are calculated from the relation $E_b=E_{Total}-E_{WS_2}-E_{h\text{-}VN}$, where $E_{Total}$, $E_{WS_2}$ and $E_{h\text{-}VN}$ are the total energies of WS$_2$/h-VN heterostructure, pristine WS$_2$ and h-VN monolayers, respectively. The calculated binding energies are -0.79 eV, -0.21 eV, -0.74 eV, -0.22 eV, -0.29 eV, and -0.31 eV, respectively, increased with the increasing interlayer spacing. The configuration in Fig. 1(a) with the lowest energy is thermodynamic preferred, in which the S atoms are directly above the V atoms and the W atoms are directly above the N atoms. In the following studies, only the most stable configuration is considered.

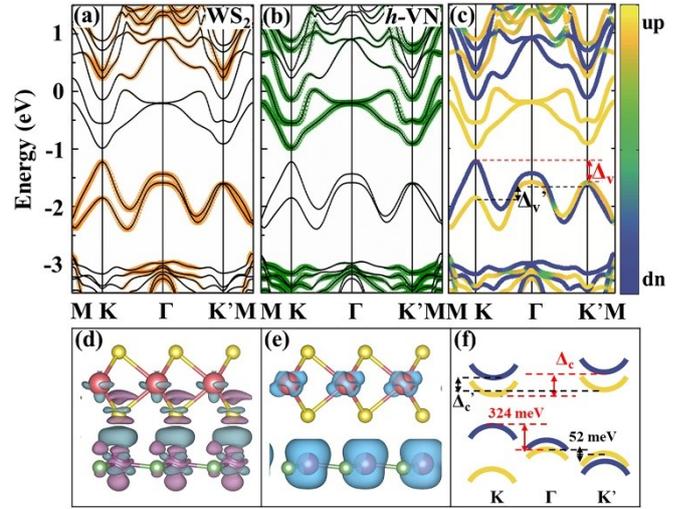

**Fig. 2** (a), (b) Atomic orbitals-projected band structures and (c) Spin-projected band structure of the WS$_2$/h-VN heterostructure. The magnitudes of the valley splitting in the first and second valence bands of WS$_2$ are denoted by $\Delta_v$ and $\Delta_v$' in (c). (d) Differential charge density with an isosurface value of 0.002 eÅ$^{-3}$. The olive (purple) distribution corresponds to charge accumulation (depletion). (e) Spin density with an isosurface value of 0.005 eÅ$^{-3}$. (f) Schematic of LCB and TVB of the WS$_2$ monolayer in WS$_2$/h-VN heterostructure, where the magnitudes of the valley splitting in the first and second conduction bands of WS$_2$ are denoted by Δ$_c$ and Δ$_c$'.

Atomic orbitals-projected band structures of the WS$_2$/*h*-VN heterostructure are shown in Fig. 2(a) and (b). Electronic states of bottom conduction band (LCB) and top valence band (TVB) are contributed by the orbitals of *h*-VN and WS$_2$ monolayer respectively, which forms a type-II band alignment. Spin-projected band structure in Fig. 2(c) suggests that the LCB and TVB of WS$_2$ monolayer still keep valley characteristic when forming the heterostructure. *K* and *K'* valleys exhibit noticeable different spin splitting of 627 and 2 meV, respectively, which are explicitly manipulated compared with that of the intrinsic value of WS$_2$. The asymmetric spin splitting between *K* and *K'* valleys generates considerable valley splitting of 376 meV for Δ$_v$ in the first valence band and 249 meV for Δ$_v$' in the second valence band. The valley splitting of the WS$_2$/*h*-VN heterostructure is found much larger than those have been reported in TMDs/bulk magnetic substrates (such as 214 meV for WS$_2$/MnO,[7] 117 meV for MoS$_2$/CoO,[24] 155 meV of MoTe$_2$/RbMnCl$_{3[36]}$). Except for the valence band, remarkable spin splitting of 130 meV and 141 meV are found at the *K* and *K'* valley of the LCB of WS$_2$. The antiparallel spins for LCB and TVB at *K* valley indicate the bright exciton states, while the parallel spins for LCB and TVB at *K'* valley is a signal of the existence of dark exciton states. This fact demonstrates that, the spin alignment of carriers in WS$_2$ can also be altered by constructing the WS$_2$/*h*-VN heterostructure. And together with the valley polarization, the dark excitons will become optically detectable and valley dependent, which is attractive for developing photon emitters for chiral optics and optically controlled information processing.

Manipulation of the dark excitons has never been predicted in the heterostructures of TMDs/bulk magnetic materials because of the strong orbital hybridization in their conduction bands.[7,23,24,26,36] Whereas the interface interaction in the 2D heterostructure in our system is basically the vdW interaction, which results in a moderate orbital coupling between each other, and is beneficial for tuning the spin and valley properties without disturbing the valley characteristic. The interfacial orbital hybridization can be observed in the differential charge density in Fig. 2(d) that a number of electrons accumulation in the WS$_2$/*h*-VN interface. Based on Bader analysis, 0.16 e of the *h*-VN monolayer transfers to the WS$_2$ monolayer, which shifts the Fermi level into the conduction band of the heterostructure. Spin density distribution shown in Fig. 2(e) indicates a ferromagnetic coupling within the heterostructure, and the induced magnetic moment of W atom is 0.2 $\mu_B$, which is four times of the reported 0.05 $\mu_B$ induced by MnO bulk[7]. As a result, the spin splittings at the *K* and *K'* points differ largely.

The large valley splitting can be attributed to the Zeeman effect in the WS$_2$/*h*-VN heterostructure. In order estimate the magnitude of the Zeeman field, the low-energy effective Hamiltonian based on the ***k.p*** model is constructed, which can be expressed as:[7,9]

$$H = at(\tau k_x \hat{\sigma}_x + k_y \hat{\sigma}_y) + \frac{\Delta}{2}\hat{\sigma}_z - \lambda\tau \frac{\hat{\sigma}_z - 1}{2}\hat{S}_z + \frac{\hat{\sigma}_z - 1}{2}(\hat{S}_z + \tau\alpha)B,$$

where *a*, *t*, Δ, *2λ*, *α*, and *B* are the lattice constant, effective hopping integral, band gap, SOC strength, orbital magnetic moment, and effective Zeeman magnetic field, respectively. $\hat{\sigma}$ are the Pauli matrices for the two base functions:

$$|d_{z^2}\rangle \text{ and } \frac{1}{\sqrt{2}}(|d_{x^2-y^2}\rangle + i\tau|d_{xy}\rangle).$$

Besides, τ is the valley index, which is +1 and −1 for the *K* and *K'* valleys, respectively. $\hat{S}_z$ is the spin operator, which has two eigenvalues of +1 and −1. The first three terms of the Hamiltonian describe the low-energy band dispersion of the pristine WS$_2$ monolayer, while the last term denotes the additional Zeeman energy term from the proximal *h*-VN monolayer. Based on the eigenvalues of the Hamiltonian, the format of valley splitting Δ$_v$ can be derived as 2(1+α$_v$)B$_v$, and the valley splitting of the second valence band Δ$_v$' (Fig. 2(c)) is introduced as 2(1−α$_v$)B$_v$. By fitting the intrinsic valley splitting with the first-principles calculation results ($\Delta_v$ = 376 and $\Delta_v$' = 249 meV), an orbital magnetic moment $\alpha_v$ of 0.20 and an effective Zeeman field $B_v$ of 156.25 meV are determined. Conversing one Bohr magneton to 5.78 × $10^{-5}$ eV T$^{-1}$, the effective Zeeman field is thus equal to a magnetic field of 2703 T.

It is notable that, the band dispersions at the $\Gamma$ point locate between the $K$ and $K'$ points for the TVB, the energy of up-spin at $\Gamma$ point is merely 52 meV higher than that of up-spin at $K'$ point, as marked in Fig. 2(f) Calculating the energy difference from down-spin at $K$ point to up-spin at $\Gamma$ point, it still has a large control range of 324 meV for the down-spin electrons. If change the spin orientation of $h$-VN by simply reversing the $h$-VN monolayer or by applying a magnetic field, the spin orientation of the band structure will be reversed. The energy difference between up-spin at $K'$ point and down-spin at $\Gamma$ point also gives a large control range of 324 meV for the up-spin electrons. If adjusting the Fermi level to locate in these regions by doping or other approaches, the spin- and valley-selectable carriers can be optically excited by control the exciting energy, and also can be electrically detected by spin-polarized scanning tunneling microscopy and anomalous Hall device. Similar to the valence bands, the valley splitting of LCB $\Delta_c$ and of the second conduction band $\Delta_c$' are calculated to be 148 meV and 123 meV, respectively, as shown in Fig. 2(f). The deduced orbital magnetic moment $\alpha_c$ is of 0.09 and effective Zeeman field $B_c$ is of 67.75 meV. By comparing LCB and TVB, we suggest that the different degree of effective Zeeman field $B_c$ and $B_v$ should be responsible for the different valley splitting behaviors for the LCB and TVB. In pristine WS$_2$ monolayer, the spin splitting is only 19 meV at $K$ and $K'$ valley of the LCB. The additional spin splittings induced by the Zeeman field are much larger, which results in the up-spin at both the $K$ and $K'$ valleys of the LCB. This explains that the LCB and TVB at $K$ valley are spin antiparallelly corresponding to the bright exciton states, while LCB and TVB at $K'$ valley are spin parallelly corresponding to the dark exciton states.

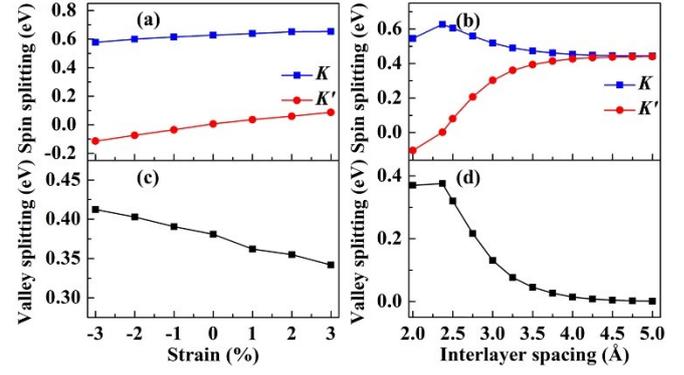

**Fig. 3** (a) Spin splitting at the $K$ and $K'$ valleys and (c) valley splitting $\Delta_v$ as a function of the in-plane strain; (b) Spin splitting at the $K$ and $K'$ valleys and (d) valley splitting $\Delta_v$ as a function of the WS$_2$/$h$-VN interlayer spacing.

Interfacial orbital interaction is closely associated with the in-plane and interlayer spacing of the heterostructure. Given this, strain engineering is expected to be a promising way to control the spin-valley splitting. By employing a various in-plane strain from -3 to 3 %, as shown in Fig. 3(a), the spin splitting at the $K$ ($K'$) valley is linearly increases from 577 to 654 meV (-114 to 87 meV, where the minus sign represents an up-spin shifting to down-spin). The large spin splitting and different responses of the $K$ and $K'$ valleys provide a large and tunable valley splitting in the region between 412 and 342 meV, as shown in Fig. 3(c). The maximum effective Zeeman field of 172.75 meV is obtained at the -3 % strain., equivalent to a magnetic field of 2989 T. Further increasing the tensile or compressive stress, strong orbital coupling at TVB occurs, which disturbs the valley characteristic and thus is not considered. Besides of the strain, interlayer spacing is also adjusted to manipulate the spin-valley splitting, as shown in Fig. 3(b) and Fig. 3(d). In the $K$ valley, the spin splitting is increased from 443 to 627 meV when the interlayer spacing reduces from 4.45 to 2.37 Å, while in the $K'$ valley it switches sign from 443 to 2 meV. The valley splitting is thus enhanced with a decreasing interlayer spacing and diminished gradually as the interlayer spacing increases, which is responsible to the short-range effect of the interfacial orbital hybridization. Further decreasing the interlayer spacing to less than the fully relaxed value, the spin and valley splitting show a decreasing trend. Hence, tunable spin-valley splitting through the external in-plane strain and out-of-plane pressure strategies is demonstrated.

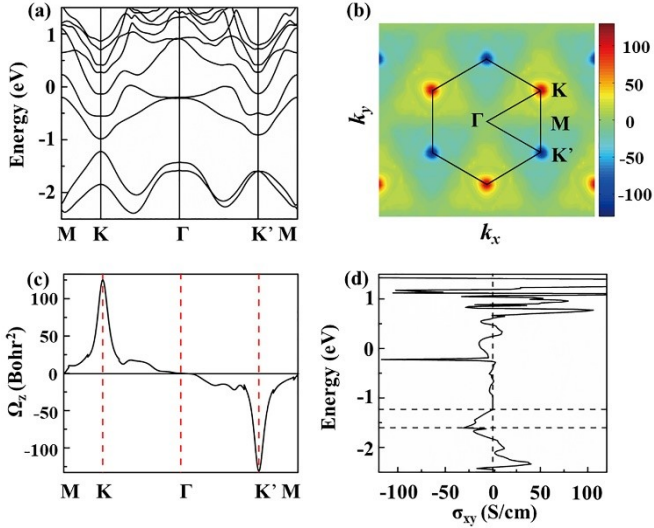

**Fig. 4** (a) Band structure of the WS$_2$/h-VN heterostructure calculated by MLWFs. Calculated Berry curvature of the WS$_2$/h-VN heterostructure (b) over 2D Brillouin zone and (c) along high symmetry lines. (d) Calculated anomalous Hall conductivity of the WS$_2$/h-VN heterostructure and the two dashed lines denote the two valley extrema.

Due to the intrinsic inversion symmetry breaking in the WS$_2$/h-VN heterostructure, the charge carriers in the K and K' valleys will acquire a nonzero Berry curvature along the out-of-plane direction. As derived from the Kudo formula, the Berry curvature can be written as a summation of all occupied contributions:[37,38]

$$\Omega_z(k) = -\sum_n \sum_{n \neq n'} f_n \frac{2Im\langle\psi_{nk}|\upsilon_x|\psi_{n'k}\rangle\langle\psi_{n'k}|\upsilon_y|\psi_{nk}\rangle}{(E_n - E_{n'})^2},$$

where $f_n$ is the Fermi-Dirac distribution function and $\upsilon_{x(y)}$ is the velocity operator; $\psi_{nk}$ is the Bloch wave function with eigenvalue $E_n$. The integral of the Berry curvature over the Brillouin zone gives the contribution to the anomalous Hall conductivity, which can be expressed as:[39]

$$\sigma_{xy} = -\frac{e^2}{\hbar} \int_{BZ} \frac{d^2k}{(2\pi)^2} \Omega_z(k).$$

A net charge current will be produced because the Hall currents from the two valleys do not completely cancel. A practical computational method for Berry curvature and Hall conductivity has been developed, namely, the Wannier interpolation, which is based on the well-constructed MLWFs.[37] To ensure the calculation accuracy of the Wannier base functions, the tight binding band structure using the MLWFs is calculated and shown in Fig. 4(a). The band dispersion is found coincided well with the DFT result [Fig. 2], which indicates that the produced Wannier base functions are sufficiently localized and the accuracy of the calculation is ensured.

Calculated Berry curvatures in the 2D Brillouin zone and along the high symmetry lines are shown in Fig. 4(b) and Fig. 4(c), respectively, with the Fermi level locating inside the band gap. The Berry curvatures for the K and K' valleys have opposite signs and slightly different absolute values, which suggests that the valley-contrasting characteristic of the WS$_2$/h-VN heterostructure is still remained. A nonzero anomalous Hall conductivity occurs owing to the broken time-reversal symmetry, as seen in Fig. 4(d).

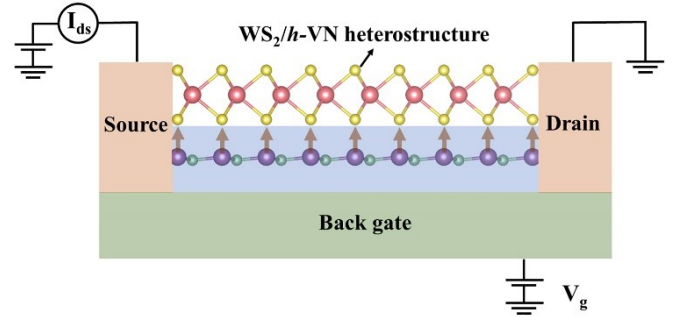

**Fig. 5** Schematic of WS$_2$/h-VN heterostructure for valleytronic devices, where the arrows denote spins.

Based on the calculated results of the Berry curvature and the anomalous Hall conductivity for the WS$_2$/h-VN heterostructure, a valleytronic device could be constructed, as shown in Fig. 5. The device is expected to realize the valley-polarized anomalous Hall effect and filter carriers with selected spin and valley indexes. By doping or other approaches, the Fermi level of WS$_2$/h-VN heterostructure can be tuned between the K valley and Γ point of the TVB

[Fig. 2(f)]. When an in-plane electric field $E$ is applied, an anomalous transverse velocity $v_\perp$ will arise due to the existence of the Berry curvature: $v_\perp \sim E \times \Omega_z(k)$. In this way, the down-spin electrons would be screened out. By reversing the $h$-VN monolayer or by applying a magnetic field, the spin orientation of $h$-VN can be changed, consequently, the up-spin electrons would also be screened out. The accumulated electrons will result in a net measurable voltage along the transversal direction, and hence, can be detected experimentally by a voltmeter. Since both the valley and spin are fully polarized, this device can also be used as a spin filter to filter all of the carriers with up- or down- spin to move transversely, which will generate a Hall current. In a word, the WS$_2$/$h$-VN heterostructure is potentially be applied to valleytronic devices for the anomalous Hall effect as both spin and valley filter, where the transport carriers move in a horizontal plane by adding an in-plane longitudinal electric field.

## IV. CONCLUSION

In this work, we study the control of valley degree of freedom in magnetic WS$_2$/$h$-VN vdW heterostructure through the first-principles calculations. By comparing the binding energies, the most stable stacking configuration of the WS$_2$/$h$-VN heterostructure is determined. Strongly asymmetric spin splitting of 627 and 2 meV are obtained respectively at $K$ and $K'$ valleys, which give rises to a large valley splitting of 376 meV in the valence band. The spin-valley splitting is induced by the orbital hybridization of the vdW interface. The effective Zeeman magnetic field is 156.25 meV, equaled to a large magnetic field of 2703 T. Considerable spin splitting is also found in the conduction band of WS$_2$, and optically detectable dark exciton states are present at the $K'$ valley. In-plane strain and interlayer spacing are employed to manipulate the strength of spin and valley splitting, to the maximum values of 654 and 412 meV, respectively. The effective Zeeman magnetic field of 2989T at most is finally achieved. The calculated Berry curvature possess the same magnitude but opposite signs for at $K$ and $K'$ valleys, which predicts opposite transverse velocities of the carriers with application of an in-plane longitudinal electric field. A nonzero anomalous Hall conductivity is further demonstrated, and a valley and spin filter device based on the WS$_2$/$h$-VN heterostructure is proposed.

## ACKNOWLEDGMENTS

We thank Penggang Li for his helps with the calculations and data analysis. The work was supported by the National Key Research and Development Program of China (Grant No. 2018YFB0406603), the National Natural Science Foundations of China (Grant Nos. 61874092, 61674124, 11604275, 61774128, 61704040, and 61804129), the Natural Science Foundation of Fujian Province of China (Grant Nos. 2018I0017 and 2017J01012), Outstanding Youth Foundation Project of Jiujiang (Grant No. 2018042), and Fundamental Research Funds for the Central Universities (Grant Nos. 20720190055 and 20720190058).